\documentclass[12pt]{article}
\usepackage{amsmath,amssymb}
\usepackage{graphicx,color}
\numberwithin{equation}{section}
\usepackage{cite}
\usepackage{bm}
\usepackage{dcolumn}  
\usepackage{amsfonts}
\usepackage[mathscr]{eucal}
\def\be{\begin{equation}} \def\ee{\end{equation}}
\def\bea{\begin{eqnarray}} \def\eea{\end{eqnarray}}
\def\bs{\boldsymbol}
\def\lb{\label} \def\mr{\mathscr{R}}
\def\a{\alpha}
 \def\b{\beta}

\newcommand{\beqar}{\begin{eqnarray*}}
\newcommand{\eeqar}{\end{eqnarray*}}

\newcommand{\z}{\zeta}

\newcommand\m{{\overline m}}
\begin{document}
\baselineskip 18pt%
\begin{titlepage}
\vspace*{1mm}%
\hfill%
\vspace*{15mm}%
\hfill
\vbox{
    \halign{#\hfil         \cr
          } 
      }  
\vspace*{20mm}

\centerline{{\Large {\bf Charged Black Holes in New Massive Gravity }}}
\centerline{{\Large{\bf    }}}
\vspace*{5mm}
\begin{center}
{ Ahmad Ghodsi\footnote{ahmad@ipm.ir, a-ghodsi@um.ac.ir} and Mohammad Moghadassi\footnote{m.moghadassi@stu-mail.um.ac.ir}}\\
\vspace*{0.2cm}
{ Department of Physics, Ferdowsi University of Mashhad, \\
P.O. Box 1436, Mashhad, Iran}\\
\vspace*{0.1cm}
\end{center}

\begin{abstract} 
We construct charged black hole solutions to three-dimensional New Massive Gravity (NMG), by adding electromagnetic Maxwell and Chern-Simons actions. We find charged black holes in the form of warped $AdS_3$ and ``log" solutions in specific critical point. The entropy, mass and angular momentum of these black holes are computed. 
\end{abstract} 

\end{titlepage}

\section{Introduction}
The three dimensional gravity has different black hole solutions. 
The BTZ black holes were found in \cite{Banados:1992wn} describe three dimensional solutions with a negative cosmological constant. Adding the higher derivative terms changes the solutions and their asymptotic behaviors.
The topological massive gravity (TMG) describes propagation of the massive gravitons around the flat, de Sitter or anti de Sitter background metrics. This theory is constructed by adding a parity-violating Chern-Simons term to the Einstein-Hilbert action \cite{Deser:1982vy}. The cosmological TMG solutions contain either the BTZ black holes \cite{Banados:1992wn} or the warped $AdS_3$ black holes \cite{Moussa:2003fc} and
\cite{Moussa:2008sj}. The black hole solutions for topologically massive gravitoelectrodynamics (TMGE) presented in \cite{Moussa:2008sj}. In this case the three-dimensional Einstein-Maxwell theory was investigated by  including both gravitational and electromagnetic Chern-Simons terms.

The new massive gravity (NMG) was found in \cite{Bergshoeff:2009hq}. This theory was constructed by adding a  parity-preserving higher derivative terms to the tree level action. In NMG theory there is also the BTZ solution as well as the warped $AdS_3$ solutions \cite{Clement:2009gq}. 
In \cite{Clement:2009ka} (and in \cite{Garbarz:2008qn} for TMG) it is obtained that for a critical coupling in NMG it is possible to find a family of massive logarithmic black holes (asymptotic to the extremal BTZ black holes in the
sense of log gravity).
For more recent works on new massive gravity see \cite{Bergshoeff:2009aq}.

In this paper and in section two, we add electromagnetic Maxwell and Chern Simons term to the new massive gravity action. Then by using the dimensional reduction procedure presented in \cite{Moussa:2008sj}, \cite{Clement:2009gq} and \cite{Clement:2009ka}, we find the equations of motion for gravitational and electromagnetic fields. In section three we use the polynomial (up to second degree) ansatz and solve the equations of motion. We will find a charged black hole in addition to the known NMG solutions. Then we will find the entropy, mass and angular momentum for this charged black hole. We find the angular momentum by using the super angular momentum approach. By writing the metric in the ADM form and using the first law of black holes thermodynamics  we can read the mass. We then discuss on the domain of validity of our solution. 

In section four, we change our ansatz to find the logarithmic charged solutions. We will use the same method of section two to compute the metric and the gauge field. Solving the equations of motion gives solutions in some specific critical points. In this case, since the theory is asymptotically AdS we can use the ADT approach to compute the conserved charges. Our computation agrees exactly with the super angular momentum approach.
\section{NMGE}
We start from a three dimensional gravitational field and couple it to an electromagnetic filed through the Lagrangian $
I_{NMGE}=I_{NMG}+I_{EM}$. The first action is  the cosmological new massive gravity theory \cite{Bergshoeff:2009hq}
\be
I_{NMG}=\frac{1}{2\kappa}\int d^3x \sqrt{|g|}\Big\{{\cal R}-\frac{1}{m^2}({\cal R}_{\mu\nu}{\cal R}^{\mu\nu}-\frac38 {\cal R}^2)-2\Lambda\Big\}\,,\label{INMG}
\ee
where $\kappa=8\pi G$ and $m$ is a relative mass parameter.
The second action is the sum of the electromagnetic Maxwell and Chern-Simons actions
\be
I_{EM}=-\frac14\int d^3x \sqrt{|g|}{\cal F}_{\mu\nu}{\cal F}^{\mu\nu}+\frac{\mu}{2}\int d^3x \epsilon^{\mu\nu\rho}{\cal A}_{\mu}\partial_\nu {\cal A}_\rho\,,\label{IEM}
\ee
where $\mu$ is the Chern-Simons coupling constant.
We try to find a stationary circularly symmetric solution for this Lagrangian by using the dimensional reduction procedure, presented in \cite{Moussa:2008sj}, \cite{Clement:2009gq} and \cite{Clement:2009ka}. We use the following ansatz for the metric and the gauge field
\be 
ds^2=\lambda_{ab}(\rho)\,dx^a dx^b + \zeta^{-2}(\rho)R^{-2}(\rho) \,d\rho^2\,,
\qquad {\cal A} = \psi_a(\rho) \,dx^a\,,  \label{metr}
\ee
where ($a,b=0,1$) and ($x^0 = t$, $x^1 = \varphi$). The parameter $\lambda$ can be expressed as a $2 \times 2$ matrix
\be
\lambda = \left(
\begin{array}{cc}
T+X & Y \\
Y & T-X
\end{array}
\right),
\ee
and $R^2\equiv \bs X^2=-T^2+X^2+Y^2$ is the Minkowski pseudo-norm of the ``vector'' $\bs X(\rho) = (T,\,X,\,Y)$.
One may reduce the action to the form of $
I = \int d^2x \int d\rho \, L \,.$
The effective Lagrangian $L$ for NMG part is given in \cite{Clement:2009gq}. Defining the vector 
$\bs L\equiv \bs X\wedge \bs X'$,\footnote{$({\bs A}\wedge {\bs B})^i=\eta^{ij}\epsilon_{jkl}A^k B^l$.}
where prime denotes the derivative with respect to $\rho$, one can write the curvature terms in the Lagrangian as follows
\bea
{\cal R} &=& \z^2\Big[-2(RR')'+\frac12(\bs X'^2)\Big] - 2\z\z'RR'\,,\nonumber\\
{\cal R}_{\mu\nu}{\cal R}^{\mu\nu}-\frac38 {\cal R}^2 &=& \z^4\Big[\frac12(\bs L'^2) - \frac14(RR')'(\bs X'^2)+\frac5{32}(\bs X'^2)^2
\Big] \nonumber\\
&+& \z^3\z'\Big[(\bs L\cdot\bs L')-\frac14RR'(\bs X'^2)\Big] +
\z^2\z'^2\frac12(\bs L^2)\,.
\eea
The electromagnetic part for the ansatz (\ref{metr}) is given in \cite{Moussa:2008sj} 
\be 
L_{EM}=\frac{\zeta}{2} {\overline \psi}' {\bs \Sigma} \cdot {\bs X} \psi'+\frac{1}{2} \mu {\overline \psi} \psi' \,.
\ee
To find the equations of motion we first consider $\zeta$ as a general function of $\rho$ then after the variations we choose a gauge, in which, $\zeta$ will be a constant value. The equations of motion for this model are as follows:

By variation of  ${\bs X}$:
\bea
\lb{varx} 
{\bs \Xi}&=&\frac{1}{2\kappa} \Big( {\bs X}\wedge ({\bs X}\wedge{\bs X}'''')+\frac{5}{2}{\bs X}\wedge ({\bs X}'\wedge{\bs X}''')+\frac{3}{2}{\bs X}'\wedge ({\bs X}\wedge{\bs X}''')\\
&+&\frac{9}{4} {\bs X}'\wedge ({\bs X}'\wedge{\bs X}'')-\frac{1}{2} {\bs X}''\wedge ({\bs X}\wedge{\bs X}'')-\bigg( \frac{1}{8}({\bs X}')^2+\frac{m^2}{\zeta^2}\bigg){\bs X}'' \Big)+\frac{m^2}{2 \zeta^2}{\overline \psi}'{\bs \Sigma} \psi'=0\,.\nonumber
\eea
Variation relative to $\zeta$ leads to the Hamiltonian constraint:
\bea 
H &\equiv &\frac{1}{2\kappa} \Big[(\boldsymbol {X}\wedge{\bs X}')\cdot ({\bs X}\wedge{\bs X}''')-\frac{1}{2}({\bs X}\wedge{\bs X}'')^2+\frac{3}{2}({\bs X}\wedge{\bs X}')\cdot ({\bs X}'\wedge{\bs X}'')\nonumber \\
&+&\frac{1}{32}({\bs X}'^2)^2+\frac{m^2}{2\zeta^2}({\bs X}')^2+ \frac{2m^2\Lambda}{\zeta^4}\Big]+\frac{m^2}{2\zeta^2} {\overline \psi}'{\bs \Sigma}\cdot {\bs X} \psi'=0\,,
\eea
and finally variation relative to $\psi$ gives
\be \label{psip}
\big( \zeta({\bs \Sigma} \cdot {\bs X}) \psi'-\mu \psi \big)'=0\,,
\ee
so the last equation gives a constant of motion, which by a gauge transformation can be set to zero \cite{Moussa:2008sj}. One may conclude from (\ref{psip}) that $
\psi'=\frac{\mu}{\zeta R^2} {\bs \Sigma}\cdot {\bs X} \psi $.
From now on we will use the same method as \cite{Moussa:2008sj}. Defining a null vector field $\lb{s} {\bs S}_E=-\frac{\kappa}{2} {\overline \psi}{\bs \Sigma} \psi$, one deduces that $
{\bs S}_E'=\frac{2 \mu}{\zeta R^2} {\bs X} \wedge {\bs S}_E
$. Also it can be shown that
\be \label{2rel}
{\overline \psi}'{\bs \Sigma}\psi'=-\frac{2\mu^2}{\zeta^2 R^2 \kappa} \big[ {\bs S}_E-\frac{2}{R^2} {\bs X} ({\bs S}_E\cdot {\bs X}) \big]\,,
\quad
{\overline \psi}'{\bs \Sigma} \cdot {\bs X}\psi'=\frac{2\mu^2}{\zeta^2 R^2 \kappa} {\bs S}_E\cdot{\bs X}\,,
\ee
which can be used to simplify the equations of motion. For example the Hamiltonian becomes
\bea
\lb{hamilsx} 
& &(\boldsymbol {X}\wedge{\bs X}')\cdot ({\bs X}\wedge{\bs X}''')-\frac{1}{2}({\bs X}\wedge{\bs X}'')^2+\frac{3}{2}({\bs X}\wedge{\bs X}')\cdot ({\bs X}'\wedge{\bs X}'')+\frac{1}{32}({\bs X}'^2)^2\nonumber \\
&+&\frac{m^2}{2\zeta^2}({\bs X}')^2+ \frac{2m^2\Lambda}{\zeta^4}+\frac{2\mu^2 m^2}{\zeta^4 R^2} {\bs S}_E\cdot {\bs X}=0\,.
\eea
\section{Black hole solution}
In this section we try to find the black hole solutions by choosing the following ansatz for the vector field ${\bs X}$
\be  
{\bs X}={\bs \alpha}\rho^2+{\bs \beta}\rho+{\bs \gamma}\,. \label{X}
\ee
Inserting this ansatz into the Hamiltonian constraint, one finds that we should insert ${\bs \alpha}^2={\bs \alpha} \cdot {\bs \beta}=0$, which is equivalent to ${\bs \alpha} \wedge {\bs \beta}=b {\bs \alpha}$. For convenience we set ${\bs \alpha}\cdot{\bs \gamma}=-z$.  Since we have the same situation as \cite{Moussa:2008sj} we choose  $\mu=\zeta$. Using the equations (\ref{2rel}) and (\ref{varx}) we are able to find
the value of ${\bs S}_E\cdot{\bs X}$ as
\bea 
\frac{2m^2}{\zeta^2 R^2}{\bs S}_E\cdot{\bs X}=(\frac{17}{4}b^2-\frac{2m^2}{\zeta^2})z+2z^2\,,
\eea
where $b^2={\bs \beta}^2$.
Using (\ref{2rel}) together with (\ref{varx}) and (\ref{hamilsx}) we find the value of ${\bs S}_E$ as
\bea\lb{s1} 
{\bs S}_E=\frac{\zeta^2}{2m^2} (R^2{\bs\alpha}+2z{\bs X})(2z+\frac{17}{4}b^2-\frac{2m^2}{\zeta^2})\,.
\eea
Putting ${\bs S}_E$ into the Hamiltonian constraint (\ref{hamilsx}) we find the following algebraic relation 
\be \lb{hsimp} 
\Big( \frac{5}{4}b^2-\frac{2m^2}{\zeta^2} \Big)z+\Big( \frac{1}{32}b^4+\frac{2m^2\Lambda}{\zeta^4}+\frac{m^2}{2\zeta^2}b^2 \Big)=0\,,
\ee
where the coefficient of $z^2$ from the gravity part has been canceled by the gauge field part. 
This is our first equation coming from the Hamiltonian constraint. Another equation will be found by the constraint $
{\bs S}_E'=\frac{2}{ R^2} {\bs X} \wedge {\bs S}_E
$, where computing ${\bs X}\wedge \mbox{(\ref{s1})}$ gives us the right hand side of it as
\be \lb{cons1}
\frac{2}{R^2}{\bs X}\wedge{\bs S}_E=-\frac{\zeta^2}{m^2} \Big( \frac{17}{4}b^2-\frac{2m^2}{\zeta^2}+2z \Big) \big(b\rho{\bs \alpha}+{\bs \alpha}\wedge{\bs \gamma}\big)\,,
\ee
and computing the derivative of (\ref{s1}) gives the left hand side of $
{\bs S}_E'=\frac{2}{ R^2} {\bs X} \wedge {\bs S}_E
$ 
\bea\lb{cons2} 
{\bs S}_E'=\frac{\zeta^2}{2m^2} \Big(\frac{17}{4}b^2-\frac{2m^2}{\zeta^2}+2z \Big)\Big\{{\bs\alpha}\Big( 2b^2 \rho-4z\rho+2{\bs \beta}\cdot{\bs \gamma} \Big)+2z \big( 2\rho{\bs \alpha}+{\bs \beta} \big)\Big\}\,,
\eea
where in the above equation we have used $R^2={\bs X}^2=b^2 \rho^2+{\bs \gamma}^2+2({\bs \alpha}\cdot{\bs \gamma})\rho^2+2({\bs \beta}\cdot {\bs \gamma})\rho$.
Comparing the equations (\ref{cons1}) and (\ref{cons2}), leads to 
\be
(b+1) ( \frac{17}{4}b^2-{2\m^2}+2z )=0\,,
\ee
where ${\overline m}^2=m^2/\zeta^2$.
This equation has two obvious solutions
$b=-1$ and $z={\overline m}^2-\frac{17}{8}b^2$, where the second one has been described in \cite{Clement:2009gq}.
\subsection{The case of $b=-1$}
By defining ${\overline \Lambda}=\Lambda/\zeta^2$, the Hamiltonian constraint in equation (\ref{hsimp}) gives
\be 
z=-\frac{1+16{\overline m}^2(1+4{\overline \Lambda})}{8(5-8{\overline m}^2)}\,.
\ee
Following to \cite{Clement:2009gq} we can choose a rotating frame and a length-time scale such that 
\be
{\bs \alpha}=(\frac12,-\frac12,0)\,,\quad
{\bs \beta}=(\omega,-\omega,-1)\,,\quad {\bs \gamma}=(z+u,z-u,-2\omega z)\,,
\ee
where $u=\frac{ \b_0^2 \rho_0^2}{4z}+\omega^2z$. By these parameters
one finds \cite{Clement:2009gq}
\be 
R^2=(1-2z)\rho^2+{\bs \gamma}^2 \equiv  \b_0^2(\rho^2- \rho_0^2)\,.
\ee
Knowing the above parameters, we are able to write the metric as warped $AdS_3$ form \cite{Clement:2009gq}
\be 
ds^2=(1- \b_0^2)\Big[dt-\Big(\frac{\rho}{1- \b_0^2}+\omega\Big)\,
d\phi\Big]^2
- \frac{ \b_0^2}{1- \b_0^2}(\rho^2- \rho_0^2)\,d\phi^2 +\frac1{ \b_0^2\zeta^2}\frac{d\rho^2}{\rho^2- \rho_0^2}\,.
\ee
One may write the metric in the ADM form by choosing $r^2 = \rho^2 +2\omega\rho +\omega^2\,(1- \b_0^2) +
\frac{ \b_0^2 \rho_0^2}{1- \b_0^2}$
\be\lb{admmetr} 
ds^2 = - \b_0^2\frac{\rho^2- \rho_0^2}{r^2}\,dt^2 + r^2\Big[d\phi
  - \frac{\rho+(1- \b_0^2)\omega}{r^2}\,dt\Big]^2 + \frac1{ \b_0^2\zeta^2}\frac{d\rho^2}{\rho^2- \rho_0^2}\,,
\ee 
where $\rho_0$ (describes the location of the horizon) together with $\omega$ are two parameters of theory. The value of $\beta_0$ is given by 
\be
\beta_0^2=\frac{21+16\m^2(-1+4\overline{\Lambda})}{4(5-8\m^2)}\,.
\ee

It remains to find the electromagnetic field for this solution. By writing ${\overline \psi}=\psi^T \Sigma^0=\left(
\begin{array}{lr}\!\! -\psi_1\,,&\!\!\!\psi_0\!\! \end{array}
\right) $, the components of ${\bs S}_E$ will be
\bea \label{efield}
S_E^0&=&\frac{\kappa}{2}(\psi_0^2+\psi_1^2)= \frac{1}{2\m^2}\big( \frac{17}{4}-2{\overline m}^2+2z \big)\Big(\frac{ \b_0^2}{2} (\rho^2- \rho_0^2) +2z \big(\frac{\rho^2}{2}+\omega \rho+u+z \big)\Big)\,, \nonumber\\
S_E^1&=&\frac{\kappa}{2}(\psi_0^2-\psi_1^2)= \frac{1}{2\m^2}\big( \frac{17}{4}-2{\overline m}^2+2z \big)\Big(-\frac{ \b_0^2}{2} (\rho^2- \rho_0^2) +2z (-\frac{\rho^2}{2}-\omega \rho-u+z) \Big)\,, \nonumber\\
S_E^2&=&\frac{\kappa}{2} (2 \psi_0 \psi_1)=- \frac{1}{\m^2} \big(\frac{17}{4}-2{\overline m}^2+2z \big) \big(\rho+2z\omega \big)z\,, 
\eea
so the first two components give
\be
\psi_0=\pm \sqrt{\frac{21-8{\overline m}^2-4 \b_0^2}{8\kappa\m^2}}\ (1- \b_0^2)\,,
\psi_1=\pm \sqrt{\frac{21-8{\overline m}^2-4 \b_0^2}{8\kappa\m^2}}\ [\rho+\omega(1- \b_0^2)]\,.
\ee
The equation for $S_E^2$ shows that the signs of the solutions must be opposite, i.e.,
\be 
{\cal A}=\psi_a dx^a= \pm \sqrt{\frac{21-8{\overline m}^2-4 \b_0^2}{8\kappa\m^2}} \Big\{ (1- \b_0^2)dt-[\rho+\omega(1- \b_0^2)] d\phi \Big\}\,.
\ee
\subsection{Entropy, mass and angular momentum}
We begin this section by computing the entropy. According to the Wald's formula the entropy for our  metric is given by \cite{Clement:2009gq}
\bea 
S= 4\pi A_h(\frac{\delta{\cal L}}
{\delta {\cal R}_{0202}}(g^{00}g^{22})^{-1})_h
= \frac{A_h}{4G}(1 -
\frac1{m^2}[(g^{00})^{-1}{\cal R}^{00} + g_{22}{\cal R}^{22} -
\frac34 {\cal R}]_h) \,,
\eea
where $A_h$ denotes the area of the horizon. We find the entropy as
\bea \label{EAH}
S=
\frac{A_h}{4G}(1 +
\frac1{2{\overline m}^2}[({\bs X}\cdot{\bs X}'') - \frac14({\bs X}'^2)])
=\frac{A_h}{8G} [\frac{8{\overline m}^2-8z-1}{4{\overline m}^2}]\,,
\eea
where similar to NMG case, here again the value of entropy is independent of $\rho_0$ and $\omega$ and it is a renormalized form of the Bekenstein-Hawking entropy by a constant factor.

On the other hand, the computation of mass and angular momentum is possible if we could linearize the field equations and use ADT approach \cite{Abbott:1981ff} and \cite{Deser:2002rt}. But instead of this, we follow Cl\'{e}ment's guess in \cite{Clement:2009gq} and compute the global charges for gravitational part by using the super-angular momentum vector. 
The super angular momentum contains two parts. The first part is coming from the NMG part of the model and is given by \cite{Clement:2009gq}
\bea \label{JNMG}
{\bs J}_{NMG}&=&-\frac{\zeta^2}{m^2}\Big\{({\bs X})^2[{\bs X} \wedge {\bs X}'''-{\bs X}' \wedge {\bs X}'']+2({\bs X} \cdot {\bs X}'){\bs X} \wedge {\bs X}''\nonumber \\
&+&\Big[ \frac{1}{8}({\bs X}')^2-\frac{5}{2}({\bs X} \cdot {\bs X}'') \Big] {\bs X} \wedge {\bs X}'\Big\}+{\bs X} \wedge {\bs X}'\,. 
\eea
The other contribution is coming from the electromagnetic part. Knowing the fact that the Lagrangian has a $SL(2,R)$ symmetry one is able to find the electromagnetic contribution to super angular momentum as
\be\label{JEM}
\bs{J}_{EM}=\frac12\bigg(\frac{\partial L_{EM}}{\partial\psi'_0}\psi_1-\frac{\partial L_{EM}}{\partial\psi'_1}\psi_0\,,
\frac{\partial L_{EM}}{\partial\psi'_0}\psi_1+\frac{\partial L_{EM}}{\partial\psi'_1}\psi_0\,,
-\frac{\partial L_{EM}}{\partial\psi'_0}\psi_0+\frac{\partial L_{EM}}{\partial\psi'_1}\psi_1\bigg)=\frac{1}{2\kappa} \bs{S}_E\,.
\ee
Summing these two contributions we find the total super angular momentum,
\be \frac{1}{2\kappa}\bs{J}_{NMGE}=\frac{1}{2\kappa}\bs{J}_{NMG}+\frac{1}{2\kappa}\bs{S}_E\,.
\ee
Inserting the values of $\bs X$ and $\psi_0$ and $\psi_1$ we find the following vector
\be
{\bs J}_{NMGE}=\frac{1}{\m^2}(2z-1)(\m^2-z-\frac18)(\rho_0^2{\bs \a}+{\bs \b}\wedge{\bs \gamma})\,.
\ee
To find the mass and angular momentum we do the same steps as \cite{Clement:2009gq}. The mass and angular momentum can be computed from
\be\label{mj}
M=-\frac{\zeta}{8G}(\delta \bs J^Y+\Delta)\,,\quad J=\frac{\zeta}{8G}(\delta \bs J^T-\delta \bs J^X)\,,
\ee
where $\delta \bs J$ is the difference between the values of the super-angular momentum for the black hole solution and for the background $( \rho_0=\omega=0)$ solution. As we will see, the value of $\Delta$ is not important for us, (it depends on the specific theory we are looking).

Inserting the values $ \rho_0=\omega=0$ in (\ref{admmetr}) the horizon-less background metric will be
\be 
ds^2=- \b_0^2dt^2+\rho^2 \big(d\phi-\frac{1}{\rho}dt\big)^2+\frac{1}{ \b_0^2 \zeta^2}\frac{d\rho^2}{\rho^2}\,,
\ee
with $
 \bar{{\bs \a}}=(\frac12,-\frac12,0)\,,  \bar{{\bs \b}}=(0,0,-1)\,, \bar{{\bs \gamma}}=(z,z,0)\,.$
Using the above relations one finds the following value for super-angular momentum in background solution
\bea 
{\bs J}_{BG}&=&-\frac{z}{\m^2}(2z-1)(\m^2-z-\frac18)\Big(1,1,0\Big)\,.
\eea
For black hole solution one finds
\bea
{\bs J}=\frac{(2z-1)}{\m^2}(\m^2-z-\frac18)\Big(\frac{\rho_0^2}{2}\big(1,-1,0\big)+ \big(u-z(1+2\omega^2),-u-z(1-2\omega^2),2\omega z\big)\Big)\,,
\eea
so the angular momentum after simplification becomes
\bea \label{J1}
J&=&\frac{\zeta}{128 G z \m^2}(8\m^2-8z-1)(2z-1)(\rho_0^2-4\omega^2 z^2)\,.
\eea
To compute the mass one can use the first law of black hole thermodynamics in the modified Smarr-like formula \cite{Moussa:2008sj} which is appropriate for warped $AdS_3$ black holes, i.e,
\be \label{fl}
M=T_HS+2\Omega_hJ\,.
\ee
Using the ADM form of the metric we can read the Hawking temperature and the horizon angular velocity as \cite{Moussa:2008sj}
\be 
T_H=\frac{1}{4\pi}\zeta r_h (N^2)'|_h\,,\quad \Omega_h=-N^\phi|_h\,.
\ee
By comparing the metrics to the following one
\be 
ds^2=-N^2 dt^2+r^2(d\phi+N^\phi dt)^2+\frac{1}{(\zeta rN)^2}d\rho^2\,,
\ee
one finds that $N^2= \b_0^2 \frac{\rho^2- \rho_0^2}{r^2}$ and $N^\phi=-\frac{\rho+(1- \b_0^2)\omega}{r^2}$
so we can read the Hawking temperature and angular velocity as
$
T_H = \frac{\zeta\b_0^2 \rho_0}{A_h}$ and $\Omega_h =
\frac{2\pi\sqrt{1-\b_0^2}}{A_h}\,,
$
where the area of the horizon is given by $A_h =
\frac{2\pi}{\sqrt{1-\b_0^2}}[\rho_0 + \omega(1-\b_0^2)]$.
Using the first law we find the following value for the mass
\bea \label{M1}
M=-\frac{\zeta}{16G\m^2}(8\m^2-8z-1)(2z-1)\omega z\,.
\eea
It is important to note that our result satisfies in differential form of first law i.e. $dM=T_H dS+\Omega_h dJ$, for both variables $\rho_0$ and $\omega$ of the black hole.
\subsection{Domain of validity}
To have a causally regular warped $AdS_3$ black hole one needs to consider $0\le\beta_0^2\le1$ or equivalently $0\le z\le \frac12$. Using this condition we can find the domain of validity of the solution
\bea\label{range1}
&&\frac{16{\overline m}^2-21}{64{\overline m}^2}\ge {\overline \Lambda}\ge -\frac{1+16{\overline m}^2}{64{\overline m}^2}\,,\quad \mbox{for} \quad {\overline m}^2\ge\frac58\,,\quad\mbox{or} \quad 0>{\overline m}^2\,,\nonumber\\
&&-\frac{1+16{\overline m}^2}{64{\overline m}^2}\ge {\overline \Lambda}\ge \frac{16{\overline m}^2-21}{64{\overline m}^2}\,,\quad \mbox{for} \quad \frac58\ge{\overline m}^2>0\,.
\eea
On the other hand the solution describes a magnetic field given by
\be
{\cal B}={\cal F}_{\rho\phi}=\mp \sqrt{{(2\kappa\m^2)}^{-1}({\frac{17}{4}-2{\overline m}^2+2z})}\,,
\ee
where the reality condition for the value of magnetic field implies $\frac{21}{8}\geq{\overline m}^2>0$.
\subsection{Relation to NMG}
As a check we can show that the NMGE solution reduces to NMG when we choose the value of $z$ as follow
\be
z={\overline m}^2-\frac{17}{8}\,.
\ee
In this value the electric field found in (\ref{efield}) will be zero and we find the same metric as  \cite{Clement:2009gq}.

The values for entropy, angular momentum and mass are given by
\be 
S=\frac{A_h}{2G{\overline m}^2}\,,\qquad
J=\frac{2z-1}{8Gz\m^2}(\rho_0^2-4\omega^2z^2)\,,\qquad
M=\frac{\zeta}{G\m^2}(1-2z)\omega z\,,
\ee
which are in agreement with \cite{Clement:2009gq}.
\section{``Log'' black hole solutions}
In this section we try to find a new type of solutions known as ``log'' solutions. We choose the following ansatz for ${\bs X}$\cite{Garbarz:2008qn}
\be
{\bs X}={\bs \a}F(\rho)+{\bs \b}G(\rho)\,.
\ee
Inserting this into the Hamiltonian constraint gives rise again to ${\bs \alpha}^2={\bs \alpha} \cdot {\bs \beta}=0$. Additionally one finds  $G=a\rho$. It also gives  the following relation for ${\bs \beta}^2=b^2$
\be \lb{ham} 
\frac{1}{32}b^4a^4+\frac{{\overline m}^2}{2}b^2a^2+2{\overline m}^2{\overline \Lambda}=0\,. 
\ee
Using the relations found in previous section one can find the value of ${\bs S}_E$ as
\be 
{\bs S}_E=\frac{1}{2m^2} a^4b^4 \zeta^{2}\rho^2{\bs \a} \left(\rho^2 F^{(4)} +4\rho F''' +\big(\frac{17}{8}-\frac{\m^2}{a^2b^2}\big) F'' \right)\,,
\ee
so the derivative of ${\bs S}$ is
\be 
{\bs S}'_E=\frac{1}{2m^2} a^4b^4\zeta^{2}\rho{\bs \a} \big( \rho^3 F^{(5)}+8\rho^2 F^{(4)}+(\frac{113}{8}-\frac{\m^2}{a^2b^2})\rho F'''+2(\frac{17}{8}-\frac{\m^2}{a^2b^2}) F'' \big)\,,
\ee
and
\be 
\frac{2}{\zeta^2 R^2}{\bs X}\wedge{\bs S}_E=-a^3b^3 \rho{\bs \a}\big( \rho^2 F^{(4)} +4\rho F''' +(\frac{17}{8}-\frac{\m^2}{a^2b^2}) F'' \big) \,.
\ee
From the constraint ${\bs S}_E\, '=\frac{2 \mu}{\zeta \mr^2} {\bs X} \wedge {\bs S}_E$ we can obtain a differential equation for $F(\rho)$ as
\be\lb{conss} 
ab\rho^3 F^{(5)} +2\big(4ab+1\big)\rho^2 F^{(4)}+ \bigg[(\frac{113}{8}-\frac{\m^2}{a^2b^2}) ab+8 \bigg]\rho F'''
+2\big(ab+1\big)\Big[\frac{17}{8}-\frac{{\overline m}^2}{a^2b^2}\Big] F''=0\,.
\ee
Since $a$ and $b$ are coming in the same footing in our equations one may fix $a$ or $b$ to one. We fix $a=1$ and obtain $b$ from equation (\ref{ham}). Without losing the generality, instead to solve directly the differential equation (\ref{conss}) we insert the following general function with arbitrary constants $A, B$ and $v$ into (\ref{conss})
\be \label{logansatz}
F(\rho)=A \rho^v\ln\rho+B\,.
\ee
In this way we find the two following algebraic equations $(l=\frac{2}{b})$ 
\bea\lb{condi1} 
\!\!\!\!\!\!\!\!\!\!\!\!&&v(v-1)(v+l)(-4v(v-1)-\frac12+\m^2l^2)=0\,,\\
\!\!\!\!\!\!\!\!\!\!\!\!&&40v^4\!+\!32(l-2)v^3+(27-48l-6\m^2 l^2)v^2+(4\m^2 l^2(1-l)+2(9 l-1))v+l(2\m^2 l^2-1)=0.\nonumber
\eea 
Solving the above equations one finds
\be\label{sols}
v=0,1\,:\,\m^2=\frac{1}{2l^2}\,,\quad v=\frac12\,:\, \m^2=-\frac{1}{2l^2}\,,\quad
v=-l\,:\,\m^2=\frac{8l^2+8l+1}{2l^2}\,.
\ee
These solutions valid when we impose (\ref{ham}). This says that they hold in critical points $l$, in terms of $\overline{\Lambda}$.
To find the metric we choose the following basis vectors as \cite{Clement:2009ka}
\be 
{\bs \a}=\frac12(1+l^2,1-l^2,-2l), \quad {\bs \b}=(1-l^{-2},-1-l^{-2},0)\,,
\ee
which leads to the metric
\be\label{ads} 
ds^2 = (-2l^{-2}\rho + F(\rho))\,dt^2 - 2lF(\rho)\,dt\,d\phi +
(2\rho + l^2F(\rho))\,d\phi^2  + \frac{l^2}{4\rho^2}d\rho^2\,. 
\ee
\subsection{Charged black holes}
Among the solutions found in (\ref{sols}) the first three cases of $v=0,\frac12, 1$ have vanishing ${\bs S}_E$ so they describe the known log-NMG solutions found in \cite{Clement:2009ka}. So here we just look at the charged black holes for the case of $v=-l$.

{\bf{The case of $ v=-l$}}:
The logarithmic charged black hole solution is given by
\bea
ds^2 &=& \Big(-2l^{-2}\rho + (A\rho^{-l}\log(\rho)+B)\Big)\,dt^2 - 2l(A\rho^{-l}\log(\rho)+B)\,dt\,d\phi\nonumber\\ 
&+& \Big(2\rho + l^2(A\rho^{-l}\log(\rho)+B)\Big)\,d\phi^2  + \frac{l^2\,d\rho^2}{4\rho^2}\,,
\eea
and
\be\label{logse}
{\bs S}_E=-16 A \frac{(l+1) (2l+1)}{l(8l^2+8l+1)} \rho^{-l} {\bs \alpha}\,.
\ee
Now we can read the gauge field as
\be \label{gaugelog}
{\cal A}=\psi_a dx^a=\pm\sqrt{{-16\frac{A}{{\kappa}}  \frac{(l+1) (2l+1)}{l(8l^2+8l+1)}\rho^{-l}}}(dt-l d\phi)\,.
\ee
The similar results is obtained for TMGE in \cite{Garbarz:2008qn}, (see equations (51) and (52)).
The above solution shows that the charged black hole found here contains both electric and magnetic fields. In order to have a real
electromagnetic field we need one of the two following conditions
\bea\label{logval}
&&1.\,\, A>0\,; \quad l<-1 \quad \mbox{or} \quad -\frac12-\frac{\sqrt{2}}{4}<l<-\frac12\,,\quad \mbox{or} \quad -\frac12+\frac{\sqrt{2}}{4}<l<0\,, \nonumber\\
&&2.\,\, A<0\,; \quad -1<l<-\frac12-\frac{\sqrt{2}}{4}\,, \quad \mbox{or} \quad  -\frac12<l<-\frac12+\frac{\sqrt{2}}{4}\,,  \quad \mbox{or} \quad 0<l\,.
\eea
The entropy of this black hole can be computed from the Wald formula evaluated at the horizon $\rho=0$ \cite{Clement:2009ka}
\be
S=\frac{\pi}{2G}\Big(\Big[(1-\frac{1}{2\m^2 l^2})\Big(2\rho+l^2 F(\rho)\Big)^\frac12\Big]-\frac{2}{\m^2}\Big[{\rho^2}{\Big(2\rho+l^2 F(\rho)\Big)^{-\frac12}}F''(\rho)\Big]\Big)_h\,.
\ee
The entropy has a finite value in the case where $l<0$. So we find
\be
S=\frac{\pi}{4\m^2 G}\sqrt{\frac{B}{l^2}}(2\m^2l^2-1)=\frac{A_h}{8\m^2 l^2 G}(2\m^2l^2-1)\,.
\ee
Again in this case the entropy is proportional to the area of the horizon.
To compute the angular momentum we can use two methods. In first method we use the information in previous section. The super-angular momentum for the NMG part is given by
\bea 
{\bs J}_{NMG}=-\frac{b^3}{\m^2}{\bs \a}\Big(-\rho^2 \Big(\rho F'''+F''\Big)+\frac18 \Big(F-\rho F'\Big)\Big)+b{\bs \a}\Big(F-\rho F'\Big)\,.
\eea
The background corresponds to $A=B=0$.
Using the relation (\ref{logse}) and by (\ref{JNMG}), (\ref{JEM}) and (\ref{mj}) one finds the following value for angular momentum
\be\label{jlog}
J=\frac{2\zeta l^2 (l+1)}{G\big(8l^2+8l+1\big)}B\,.
\ee
As an alternative way we can apply the second method. The metric in (\ref{ads}) is asymptotically AdS if  $\lim_{\rho\rightarrow\infty}(\rho^{-1}F(\rho))=0$ (or $l>-1$ in this case) so we can use the ADT approach to compute the conserved charges. The net Killing charge is given by
\be
Q(\xi)=Q_G(\xi)+Q_{EM}(\xi)\,.
\ee  
The $Q_G$ is the Killing charge corresponding to the gravitational part and is given by \cite{Clement:2009ka}
\be
Q_G(\xi)=\frac{\zeta}{G\m^2 l^3}\Big(\rho^3 F'''+\rho^2 F''+\frac14 (\frac12-\m^2l^2)(\rho F'-F)\Big)([{\bs\a}]\xi)^0\,,
\ee
where $[{\bs\a}]$ is the matrix
\be
{\bs\a}=\left(
\begin{array}{cc}
-\a^Y & -\a^T+\a^X \\
\a^T+\a^X & \a^Y
\end{array}
\right).
\ee
To find the angular momentum we need the Killing vector $\xi=(0,1)$. Then we find
\be\label{QG}
Q_G(\xi)=\frac{\zeta}{G\m^2}l^{-1}(l+1)\Big((2l+1) A\rho^{-l}+l B\Big)\,.
\ee
In addition for the background $(A=B=0)$ we have
$
Q^{BG}_{G}=0\,.
$
The $Q_{EM}$ is the Killing charge with respect to the electromagnetic part. The difference between the solution and the background in electromagnetic part can be found in \cite{Moussa:2003fc}
\be
\delta Q_{EM}(\xi)=\frac{\zeta\pi}{m^2\kappa}\Big\{\xi^T {\bs \Sigma}.({\bs S}_E-{\bs S}^{BG}_E)-\kappa(\bar{\psi}-\bar{\psi}^{BG})\psi^{BG}\xi^T\Big\}^0\,.
\ee
In this case the background $A=B=0$ results the vanishing of the gauge field so we find
\be\label{QEM}
\delta Q_{EM}(\xi)=-\frac{\zeta}{G\m^2}l^{-1}(l+1)(2l+1) A\rho^{-l}\,.
\ee
Then the angular momentum is
\be\label{J2}
J=\delta Q_{G}(\xi)+\delta Q_{EM}(\xi)=\frac{2\zeta l^2 (l+1)}{G\big(8l^2+8l+1\big)}B\,,
\ee
where it is exactly equal to (\ref{jlog}) and independent of $\rho$.

To find the mass of the solution we can use the second method and insert the Killing time like vector $\xi=(-1,0)$. It simply gives
$M=\frac{J}{l}\,.$
We can find the Hawking temperature for this black hole, which is equal to zero, $
T_H=\frac{1}{2\pi}\sqrt{g^{\rho\rho}}\frac{d}{d\rho}\sqrt{g_{tt}}|_h=0\,.$
\section{Summary and Discussion}
In this paper we find the charged solution for the three dimensional new massive gravity by adding the Maxwell and Chern-Simons term into the action. We use the reduced action approach to find the equation of motion for gravitational and gauge fields. Then we consider two types of ansatz. In first case we consider the second order polynomial solution (\ref{X}). Our results contain those solutions with vanishing gauge field (uncharged) which has been found previously and a charged black hole solution in warped $AdS_3$ form. 

We show that the entropy of such a black hole (\ref{EAH}), is similar to uncharged cases and is proportional to the area of the horizon. We also find the angular momentum (\ref{J1}) and the mass (\ref{M1}) for this charged black hole using the first law of thermodynamics (\ref{fl}) for black holes. Our computation shows that this solution is valid in a special range of cosmological constant $\Lambda$ with respect to relative mass parameter $m^2$ and the Chern-Simons coupling constant $\mu=\zeta$ (\ref{range1}).

In second case we consider the logarithmic ansatz (\ref{logansatz}). Solving equations of motion gives us different solutions in different critical points. Again in this case there are uncharged solutions as well as a charged solution. The charged solution corresponds to $v=-l$ with relative mass parameter $m^2=\mu^2 (\frac{8l^2+8l+1}{2l^2})$. At this point the cosmological constant is given by 
$
\Lambda=\mu^2\frac{32l^2+32l+5}{2l^2(8l^2+8l+1)}\,.
$
The domain of validity for this black hole is when $l>-1$ together with (\ref{logval}).

Since in this type of solutions the asymptotic of the theory is the AdS space we can find the conserved charges of the theory by using the ADT formalism. We show that the infinite parts of the charges in gravitational part (\ref{QG}) and in the electromagnetic part (\ref{QEM}) cancel each other and give a finite value for mass $M=J/l$ and angular momentum (\ref{J2}) (which exactly agrees with the value found in super angular momentum approach (\ref{jlog})). 

{\bf{Acknowledgment:}}
This work was supported by Ferdowsi University of Mashhad under the grant 2/15343, (11/07/1389).

\end{document}